\documentclass[aps,prl,showpacs,preprintnumbers,twocolumn,superscriptaddress]{revtex4-2}
\usepackage{amsmath,amssymb}
\usepackage{bm}
\usepackage{tipa}
\usepackage{upgreek}
\usepackage{comment}
\usepackage{mathrsfs}
\usepackage{graphicx}
\usepackage{braket}
\usepackage{enumitem}
\usepackage{mathbbol}
\usepackage{booktabs}
\usepackage{gensymb}
\usepackage[normalem]{ulem}
\usepackage{color}
\usepackage[colorlinks,bookmarks=true,citecolor=blue,linkcolor=red,urlcolor=blue]{hyperref}
\usepackage{hyperref}

\usepackage{pifont}

\newcommand{\xmark}{\ding{55}}

\begin{document}
	
\title{Kekul\'e spirals and charge transfer cascades in twisted symmetric trilayer graphene}

	\author{Ziwei Wang}
	\affiliation{Rudolf Peierls Centre for Theoretical Physics, Parks Road, Oxford, OX1 3PU, UK}

 \author{Yves H. Kwan}
	\affiliation{Princeton Center for Theoretical Science, Princeton University, Princeton NJ 08544, USA}
 	\author{Glenn Wagner}
	\affiliation{Department of Physics, University of Zurich, Winterthurerstrasse 190, 8057 Zurich, Switzerland}
	\author{Nick Bultinck}
	\affiliation{Rudolf Peierls Centre for Theoretical Physics, Parks Road, Oxford, OX1 3PU, UK}
	\affiliation{Department of Physics, Ghent University, Krijgslaan 281, 9000 Gent, Belgium}
	\author{Steven H. Simon}
	\affiliation{Rudolf Peierls Centre for Theoretical Physics, Parks Road, Oxford, OX1 3PU, UK}
	\author{S.A. Parameswaran}
	\affiliation{Rudolf Peierls Centre for Theoretical Physics, Parks Road, Oxford, OX1 3PU, UK}

\begin{abstract}
We study the phase diagram of magic-angle twisted symmetric trilayer graphene in the presence of uniaxial heterostrain and interlayer displacement  field.  For experimentally reasonable strain, our mean-field analysis finds robust Kekul\'e spiral order whose doping-dependent ordering vector is incommensurate with the moiré superlattice, consistent with recent scanning tunneling microscopy experiments,   and paralleling the behaviour of closely-related   twisted bilayer graphene (TBG) systems. Strikingly, we  identify a new possibility absent in TBG: the existence of {\it commensurate} Kekul\'e spiral order even at zero strain for experimentally realistic values of the interlayer potential in a trilayer. Our studies also reveal a complex pattern of charge transfer between weakly- and strongly-dispersive bands in strained trilayer samples as the density is tuned by electrostatic gating,  that can be understood intuitively in terms of the `cascades' in the compressibility of magic-angle TBG.
\end{abstract}

\maketitle

\textit{Introduction.---} The discovery of superconductivity proximate to correlated insulating behaviour~\cite{Cao2018, Cao2018b, Yankowitz_2019, Lu2019, Stepanov_2020, saito_independent_2020, Arora2020SC, Cao_2021, Oh2021unconventional} in twisted bilayer graphene (TBG) in the `magic angle' regime~\cite{Bistritzer2011, tarnopolsky_origin_2019} has stimulated intense investigation of a host of other multilayer graphene systems. Among these, alternating-twist multilayers exhibit identical moir\'e patterns between each pair of adjacent layers and have well-defined magic angles in the chiral limit~\cite{khalaf_magic_2019}. The simplest member in the group, twisted symmetric trilayer graphene (TSTG), which is composed of three layers of graphene stacked together
with the inner layer rotated, has been experimentally studied~\cite{park_tunable_2021, zhang_ascendance_2021, cao_pauli-limit_2021, kim_evidence_2022, turkel_orderly_2022, zhang_electronic_2022, yang_wafer-scale_2022, hao_electric_2021, shen_dirac_2023, liu_isospin_2022, lin_zero-field_2022, li_observation_2022, kim_imaging_2023}, especially in the context of its superconducting properties, and has received extensive theoretical investigation~\cite{li_electronic_2019, carr_ultraheavy_2020, tritsaris_electronic_2020, lopez-bezanilla_electrical_2020, wu_lattice_2021, calugaru_twisted_2021, shin_stacking_2021, lei_mirror_2021, qin_-plane_2021, phong_band_2021, xie_twisted_2021, choi_dichotomy_2021, wu_magic_2021, lake_reentrant_2021, ledwith_tb_2021, fischer_unconventional_2022, guerci_higher-order_2022, scammell_theory_2022, christos_correlated_2022, xie_alternating_2022, classen_interaction-induced_2022, leconte_electronic_2022, lin_energetic_2022, samajdar_moire_2022, gonzalez_ising_2023, shin_electronic_2023, yu_magic-angle_2023}. In the absence of external fields, the single-particle Hilbert space of TSTG can be decomposed into a direct sum of monolayer graphene and TBG parts~\cite{khalaf_magic_2019}, allowing many phenomena in TBG to be reproduced in TSTG. If one applies a perpendicular electric field, the TBG and graphene sectors are hybridized, allowing for further tunability of the system.

Among the various correlated states in TBG, a novel form of translation-symmetry breaking order dubbed the incommensurate Kekul\'e spiral (IKS) has been proposed by some of us to characterize the many-body ground state at all integer fillings except charge neutrality~\cite{kwan_kekule_2021}, as well as a range of non-integer fillings~\cite{wagner_global_2021}. The signature of the proposed state, which is stabilized by experimentally realistic uniaxial strains, consists of a graphene scale Kekul\'e-like pattern with complex spatial dependence on the moiré scale, and has recently been observed with high-resolution scanning tunnelling microscopy (STM)~\cite{nuckolls2023quantum}. A similar pattern has also been observed in TSTG~\cite{kim_imaging_2023}, but  Kekul\'e spiral order in TSTG is 
theoretically unexplored, motivating the present work.

\begin{figure}
\centering
\includegraphics[width=1\linewidth]{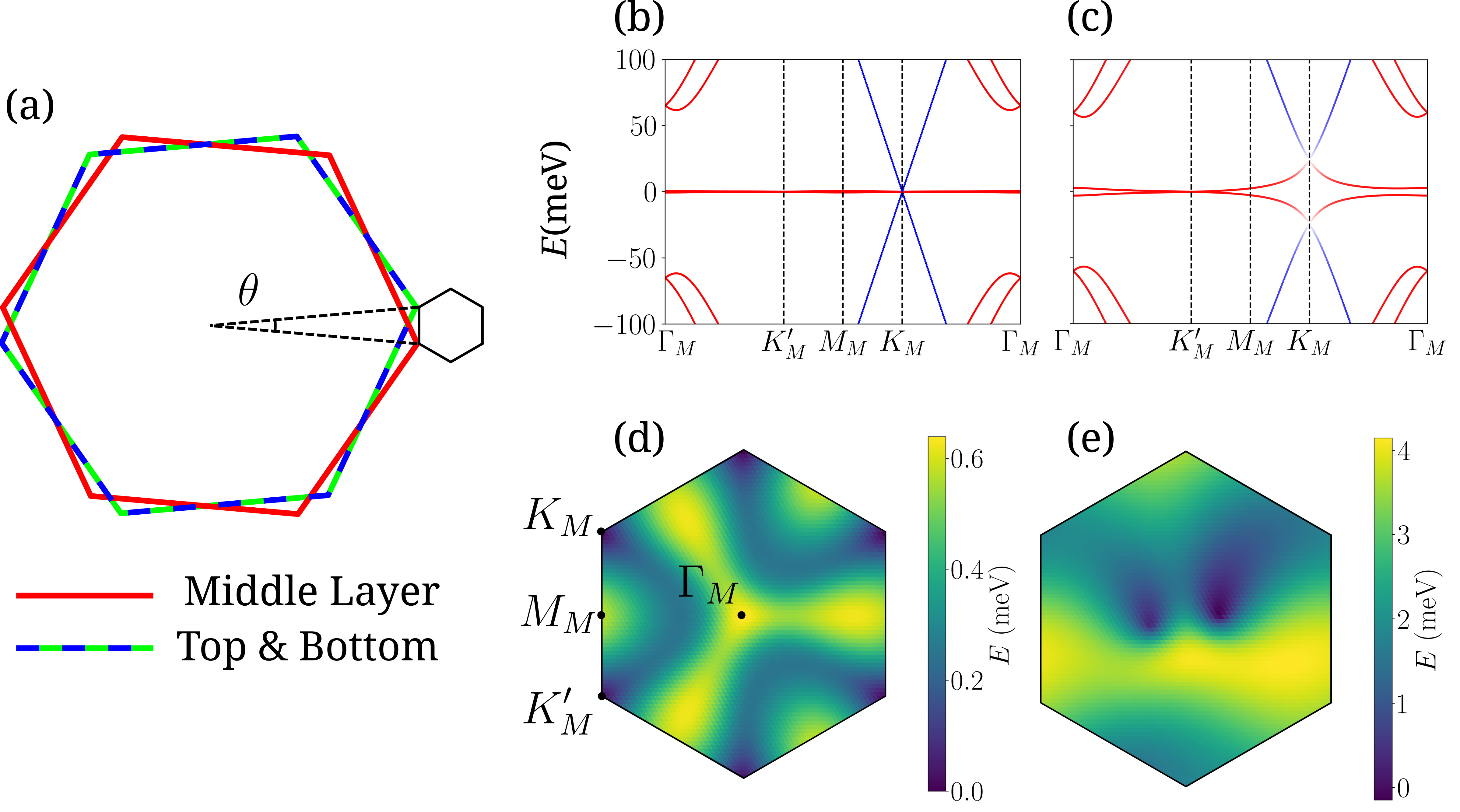}
\caption{\textbf{Twisted symmetric trilayer graphene.} (a) The large hexagons denote graphene Brillouin zones of individual layers and the small hexagon denotes the moir\'e BZ for valley $K$. 
(b) Without interlayer potential, TSTG decomposes into a TBG sector (red) and a graphene sector (blue) with opposite mirror eigenvalues. (c) A finite interlayer potential (100 meV) breaks mirror symmetry and hybridizes the two sectors. The color represents the relative weight of the wavefunction on the zero-field TBG and graphene sectors.
(d,e) TBG sector conduction band at zero interlayer potential without and with strain ($\epsilon=0.1\%$) respectively.  Strain breaks $C_3$-symmetry, unpinning the Dirac points, and increases the bandwidth. The corresponding band structures in valley $K'$ can be found using time-reversal symmetry.
} 
\label{model}
\end{figure}
Here, we perform extensive Hartree-Fock (HF) simulations of TSTG that incorporate the effect of strain and allow for translation symmetry breaking. As in the case of TBG, we find IKS to be ubiquitous for non-zero integer and non-integer fillings of TSTG at finite strain. Furthermore, we find that \textit{commensurate} Kekul\'e spiral (CKS) order in unstrained TSTG  can be accessed even at zero strain by applying a large interlayer potential --- a  possibility absent in TBG. These results should be contrasted with previous studies of  TSTG~\cite{xie_twisted_2021, ledwith_tb_2021, christos_correlated_2022, gonzalez_ising_2023, yu_magic-angle_2023} at zero strain. We also explore the process of `charge transfer' between the TBG and graphene sectors of TSTG at zero interlayer potential, and elucidate subtleties in defining commensurate fillings in light of this mechanism.

\textit{Model and Symmetries.---}TSTG is made by stacking three sheets of graphene with the middle layer twisted by an angle $\theta$ relative to the top and bottom layers [Fig.~\ref{model}(a)]. The moir\'e patterns generated within the top and bottom pairs of layers are identical, preserving the overall moir\'e periodicity of the system. The single-particle Hamiltonian of TSTG can be decomposed~\cite{khalaf_magic_2019} into a mirror-symmetry even sector, which is effectively a renormalized version of TBG, and a mirror-symmetry odd sector, which resembles monolayer graphene [Fig.~\ref{model}(b)]. The magic angles of TSTG and TBG are  related via $\theta_{\text{TSTG}} =\sqrt{2}\theta_{\text{TBG}} \approx 1.56 ^\circ$. In contrast to TBG, an electric displacement field normal to the layers has a significant effect: it produces an interlayer potential that breaks the mirror symmetry and mixes the odd and even sectors [Fig.~\ref{model}(c)]. This modifies the dispersion, expanding the parameter space for new phases.

We also incorporate the effect of strain in the system. Homostrain (where all layers experience the same strain) has only a small effect on TBG~\cite{huder_electronic_2018}, and we expect it to play a similarly minor role in TSTG. Therefore,  we restrict our attention to heterostrain, produced by unequal strain in the different layers.
%where strains of different layers are unequal.
In principle, strain could break the mirror symmetry of the system and introduce incommensurate moir\'e patterns for the top and bottom pairs of layers, potentially leading to a supermoir\'e pattern. However, we focus here on the case where top and bottom layers experience %the same 
identical strains, thus preserving the moiré periodicity of the system. This choice can be justified as STM experiments observe well defined moir\'e patterns in large regions of the TSTG samples~\cite{kim_evidence_2022, turkel_orderly_2022, kim_imaging_2023}. We also ignore the small-angle Pauli-matrix rotations %twists 
in the Dirac kinetic terms in each layer; then, absent strain or interlayer potential, %that without strain or interlayer potential, 
the system has exact particle-hole symmetry, which is lost when both perturbations are included. 

\textit{Phase Diagram.---}We perform band-projected self-consistent HF calculations on the model discussed above at the magic angle $\theta=1.56^\circ$. We use realistic hopping parameters $w_{AA} = 75$\,meV and $w_{AB} = 110$\,meV, and include strain $\epsilon$ and interlayer potential $\Delta V$ as free parameters of the model. We model interactions via the dual-gate screened Coulomb potential $V(q) = (e^2/2\epsilon_0\epsilon_rq) \tanh qd$, with a screening length $d = 25$ nm and relative permittivity $\epsilon_r = 10$. To avoid double-counting interactions, we use the ``average'' subtraction scheme~\cite{SupMat}. %detailed in the Supplementary Materials.
Despite the lack of exact particle-hole symmetry for $\Delta V,\epsilon\neq0$, we find that the phase diagrams at $\nu$ and $-\nu$ are qualitatively similar. Hence we only present results for $|\nu|\geq 0$ in the main text, and refer readers to Ref.~\cite{SupMat} for $\nu<0$ results.

\begin{figure*}[t!]
    \centering
\includegraphics[width=1\linewidth]{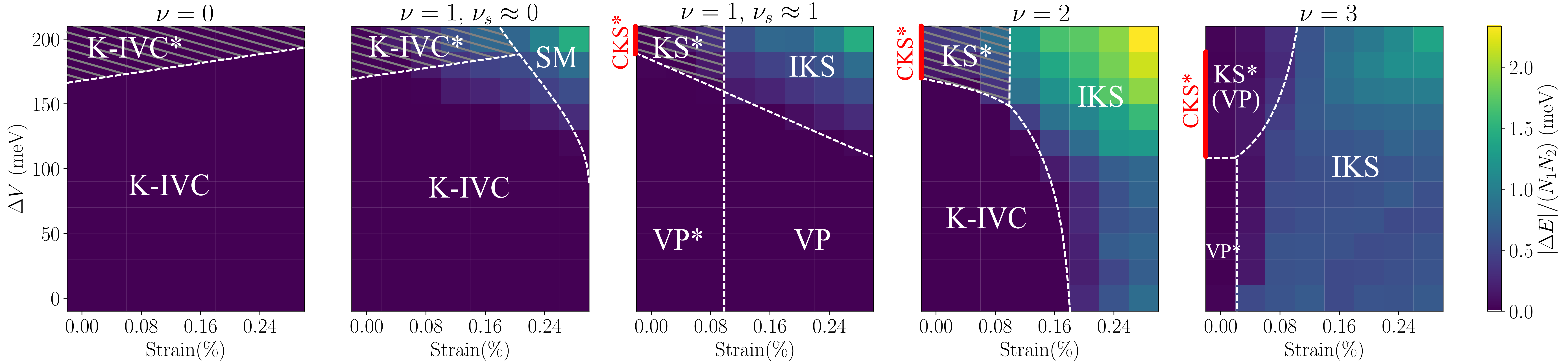}
    \caption{Phase diagrams as a function of strain and interlayer potential at integer fillings $\nu$. We have performed $10 \times 10$ self-consistent Hartree-Fock calculations, scanning over all possible Kekul\'e spiral vectors $\bm{q}$. $\Delta E$ is the energy difference between the $\bm{q} = 0$ state with the overall energy minimum. The acronyms are: K-IVC for Kramers-intervalley coherence, VP for valley polarization, (I)KS for (incommensurate) Kekul\'e spiral, SM for (compensated) semi-metal. Asterisk (*) denotes states that break $C_2T$-symmetry. The KS*(VP) region in $\nu = 3$ consists of Kekul\'e spiral states with possible finite valley polarization. The red lines on the zero-strain axis indicate \textit{commensurate} Kekul\'e spirals. The shaded regions have a finite charge gap. For $\nu = 1$, we have shown the phases and charge gaps of the two spin sectors separately, where $\nu_s$ denotes the filling of each spin sector. As $\Delta E$ cannot be separately defined for each spin sector, the data are simply duplicated in the two plots of $\nu = 1$.} 
    \label{phase_diagram}
\end{figure*}

Unless otherwise noted, we assume collinear spin configurations, but we have also checked that, for the parameter range in Fig.~\ref{phase_diagram}, relaxing this condition does not lead to different ground states up to $\text{SU}(2)_{K} \times \text{SU}(2)_{K^\prime}$ valley-dependent spin rotations. In order to capture the incommensurate Kekul\'e spirals (IKS) proposed in Ref.~\cite{kwan_kekule_2021} or its commensurate counterparts, we allow for coherence between states of momentum $\bm{k} - \tau\bm{q}/2$ in valley $\tau$ with states of momentum $\bm{k} - \tau^\prime\bm{q}/2$ in valley $\tau^\prime$ (here $\tau$ and $\tau^\prime$ take on values of $\pm 1$) for some Kekul\'e spiral vector $\bm{q}$ in our HF calculations, and then optimize over all $\bm{q}$. One could further consider the slightly more general situation where the two spin sectors have different $\bm{q}$. As such, the one-particle density matrix takes the form
\begin{equation}\label{eq:Pqs}
    \braket{\hat{c}^\dagger_{\bm{k} - \tau\bm{q}_s/2,\tau s a}\hat{c}_{\bm{k} - \tau^\prime\bm{q}_s/2, \tau^\prime s b}} = P_{\tau a; \tau^\prime b}(\bm{k}, s)
\end{equation}
where $s$ is spin index, $a$ and $b$ are single particle band indices, and $\bm{q}_s$ is the Kekul\'e spiral vector with possible spin-dependence. In the majority of cases, this spin-dependence is not relevant and we simply scan over $\bm{q}_\uparrow = \bm{q}_\downarrow \equiv
 \bm{q}$. Spin-dependence of Kekul\'e spiral vectors becomes relevant when both spin sectors have intervalley coherence and the two sectors are nonequivalent, as is the case for $\nu = 1$. As such, we have allowed $\bm{q}_\uparrow\neq\bm{q}_\downarrow$ when computing the phase diagram at $\nu=1$.  For the range of interlayer potentials considered in the main text, we found that allowing for
more generic translation symmetry breaking beyond Eq.~\ref{eq:Pqs} 
does not lead to different results. At higher interlayer potentials, we observe a complex pattern of translational symmetry breaking with nontrivial spin structure~\cite{SupMat}.

Fig.~\ref{phase_diagram} shows the phase diagrams for $\nu = 0, 1, 2, 3$. % are presented in Fig.~\ref{phase_diagram}.
Kekul\'e states involve non-zero Kekul\'e vector $\bm{q}$ and finite intervalley coherence. To diagnose the presence  of these states and demonstrate their robustness, we show energy differences per unit cell between the lowest-energy   $\bm{q} = 0$ solution  and the energy achieved by minimizing over all $\bm{q}$: non-zero values indicate Kekul\'e spiral ground states. 

At zero interlayer potential, 
the single-particle Hamiltonian decomposes into mirror-symmetry even (TBG) and odd (graphene) sectors. For all fillings and strains examined here, we observe no spontaneous breaking of this symmetry, and the two sectors are only coupled by Hartree terms. Compared to pure TBG, the inclusion of the graphene sector can lead to `charge-transfer cascades' from the TBG sector, discussed below, but otherwise the phase diagram is consistent with the corresponding one for TBG~\cite{kwan_kekule_2021}. We find  that IKS order appears at $\nu = 2, 3$ at finite strain (Fig.~\ref{phase_diagram}).  While for the range of strains shown  in Fig.~\ref{phase_diagram} we find Kramers intervalley-coherent (K-IVC) state at $\nu=0,1$, for larger strains the system transitions into IKS at $\nu = 1$ and  a symmetric state at $\nu = 0$, again consistent with the situation in TBG.
 
A non-zero interlayer potential mixes the two symmetry sectors, and the simple picture of Hartree-coupled TBG and graphene sectors is no longer applicable. The mixing of relatively flat TBG bands with graphene bands leads to more dispersive central bands. This favors the formation of IKS by reducing the minimal strain required for its stabilization, except at charge neutrality, where no Kekul\'e spiral is observed for either TBG or TSTG.

At high interlayer potential of about $200$ meV, a common feature  except at $\nu = 3$, is the breaking of $C_2T$ symmetry and the opening  of a charge gap. While strong $C_2T$-breaking has not been observed in monolayer graphene under ordinary conditions, as graphene orbitals are hybridized with TBG orbitals, the Dirac point becomes less dispersive and interaction-induced spontaneous symmetry breaking becomes possible. For $\nu = 3$, $C_2T$ symmetry only spontaneously breaks in one of the two spin sectors, leaving the system gapless. We note that Ref.~\cite{shen_dirac_2023} finds a charge gap at $\nu = 2$, in qualitative agreement with our result.

The phase diagram is most complex at $\nu = 1$, where the two spin sectors have densities of approximately 0 and 1 electron per unit cell relative to their respective charge neutrality points (finite but small charge transfers between the two sectors could occur). The two sectors are only coupled with Hartree interaction, and otherwise exhibit their own physics. To capture the physics correctly, we allow for different Kekul\'e vectors for the two sectors \footnote{For numerical efficiency we fix one spin sector to have $\bm{q} = 0$ as we observe no Kekul\'e spirals at $\nu = 0$. We have also performed HF allowing for most general symmetry breaking to confirm the adequacy of the method.}. In particular, we find ground states with K-IVC ($\bm{q}_\uparrow = 0$) in the spin sector at charge neutrality and IKS ($\bm{q}_\downarrow \neq 0$) in the other spin sector.

\textit{Commensurate Kekul\'e Spirals.---}A new phase that we identify in TSTG is the commensurate Kekul\'e spiral (CKS*)~\footnote{The asterisk emphasises $C_2T$-symmetry breaking, though in TSTG we do not find $C_2T$-conserving commensurate Kekul\'e spiral states.} at high interlayer potential and zero strain. In CKS*, the spiral wavevector relates the $K_M$-point of valley $K$ (Fig.~\ref{model}(c)) where TBG and graphene hybridization is most prominent, to the $\Gamma$-point of valley $K^\prime$. As such, the resulting spiral order is commensurate with moir\'e periodicity. We note that  CKS* further differs from IKS  in that it breaks $C_2T$ symmetry. %In the phase diagram 
This is also true of other `starred' states, including the KS*  at $\nu = 1, 2, 3$. At zero strain all KS* states have commensurate $\bm{q}$ (except $\nu = 3$ and large $\Delta V$), but this commensurability is lost once any finite strain is introduced (though the graphene scale lock-in mechanism introduced in Ref.~\cite{kim_imaging_2023} may stabilize commensuration at finite strain). While CKS* is possible at all of $\nu = 1, 2, 3$, it is only robust (i.e.~significant energy advantage over competing states) at $\nu = 2$, where it is the ground state up to about $\Delta V=240$~meV.

\textit{Charge-Transfer Cascades.---}At zero interlayer potential, the TBG and graphene sectors are coupled only by Hartree terms as the mirror symmetry is unbroken. While the total electron density is fixed, electronic charge could transfer non-trivially between the two sectors, as first noted in Ref.~\cite{xie_twisted_2021} at $\nu = 3$ and large chiral ratio.  We now track this charge-transfer mechanism across general fillings, and with finite strain.  We use a HF interpolation scheme~\cite{xie_twisted_2021} to mitigate finite-size effects on the dispersion and hence more accurately compute the charge transfer.

\begin{figure}
\centering
\includegraphics[width=1\linewidth]{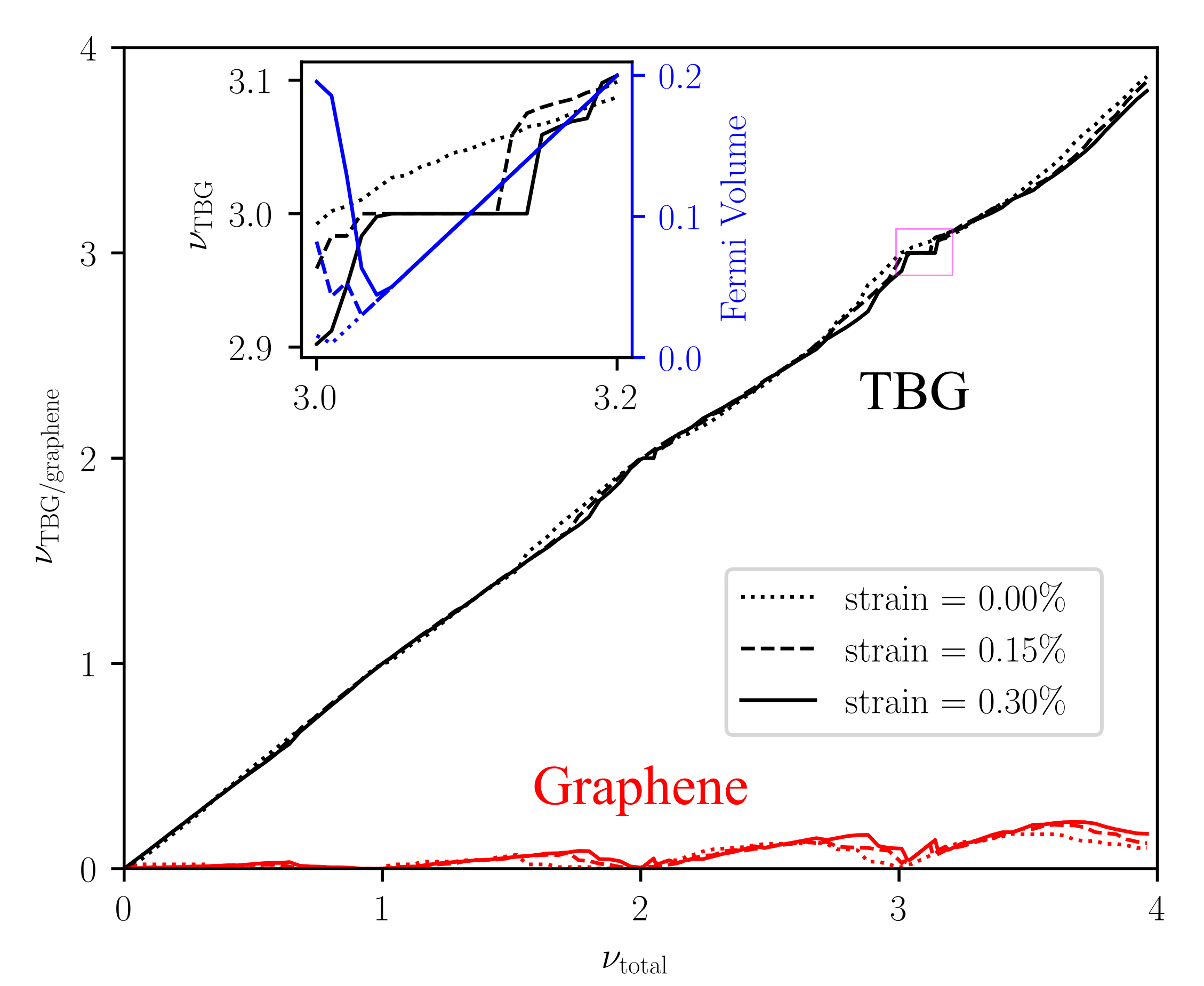}
\caption{`Charge transfer plateaus' in TSTG. We plot the fillings of TBG and graphene sectors from $10 \times 10$ self-consistent HF, interpolated to $60 \times 60$, as a function of total fillings. We notice plateaus at integer TBG fillings, which are due to a correlated charge gap in the TBG sector. In the inset, we additionally show the total unsigned Fermi volumes of all bands (`Fermi Volume') as a measure of the metallicity of the system. We notice that, due to non-trivial charge transfer between the two sectors, minima in `Fermi Volume' need not occur exactly at total integer fillings.}
\label{charge_transfer}
\end{figure}

Fig.~\ref{charge_transfer} shows the filling of the TBG sector $\nu_{\text{TBG}}$ as a function of total filling $\nu$. We observe plateaus at integer values of $\nu_{\text{TBG}}$ near $\nu = 2$ and $\nu = 3$. This can be rationalized in terms of the formation of a correlated insulator with a finite charge gap in the TBG sector, such that injection of a small amount of additional charge  only changes the filling in the graphene sector. The precise width of the plateau depends on the size of the TBG charge gap and the density of states of graphene at the chemical potential. We also find that the sum of unsigned Fermi volumes of all bands --- a rough proxy for %an indicator of 
 how `metallic' the system is --- can reach a local minimum at non-integer total fillings, meaning that the most insulator-like filling does not necessarily occur at integer $\nu$ due to the charge transfer.

In generating Fig.~\ref{charge_transfer}, we have determined the charge-transfer from self-consistent HF calculations of the full system. However, if we treat the Hartree interaction between the two sectors as a function of only the densities of the respective sectors (which can be justified by assuming the graphene sector charge density is approximately uniform), we can write
\begin{equation}
\begin{split}
E(n) & = E_{\text{TBG}}(n_{\text{TBG}}) + E_{\text{graphene}}(n - n_{\text{TBG}}) \\ 
 &\,\,\,\,  +U_0n_{\text{TBG}}(n - n_{\text{TBG}}) 
\end{split}
\end{equation}
where $n$ denotes the total electron density and $U_0 = V(\bm{q} = 0)$. We can re-write this as
\begin{equation}
    E(n) = \tilde{E}_{\text{TBG}}(n_{\text{TBG}}) + \tilde{E}_{\text{graphene}}(n - n_{\text{TBG}})
\end{equation}
where $\tilde{E}_{\text{TBG}}(n_{\text{TBG}}) = E_{\text{TBG}}(n_{\text{TBG}}) - U_0n_{\text{TBG}}^2/2$ is the energy of TBG sector with the electrostatic energy of uniform electron gas subtracted away, and similarly for $\tilde{E}_{\text{graphene}}$. We have also dropped terms that depend only on total electron density.    Minimization of the total energy corresponds to $dE/dn_{\text{TBG}} = 0$, giving
\begin{equation}
    \tilde{\mu}_{\text{TBG}} = \tilde{\mu}_{\text{graphene}}.
\end{equation}
The graphene sector can be modelled as a non-interacting Dirac cone, so that we can estimate the charge transfer from the interacting physics of TBG sector alone. The result is consistent with full TSTG HF calculations~\cite{SupMat}.

\begin{figure}
    \centering
    \includegraphics[width = 1\linewidth]{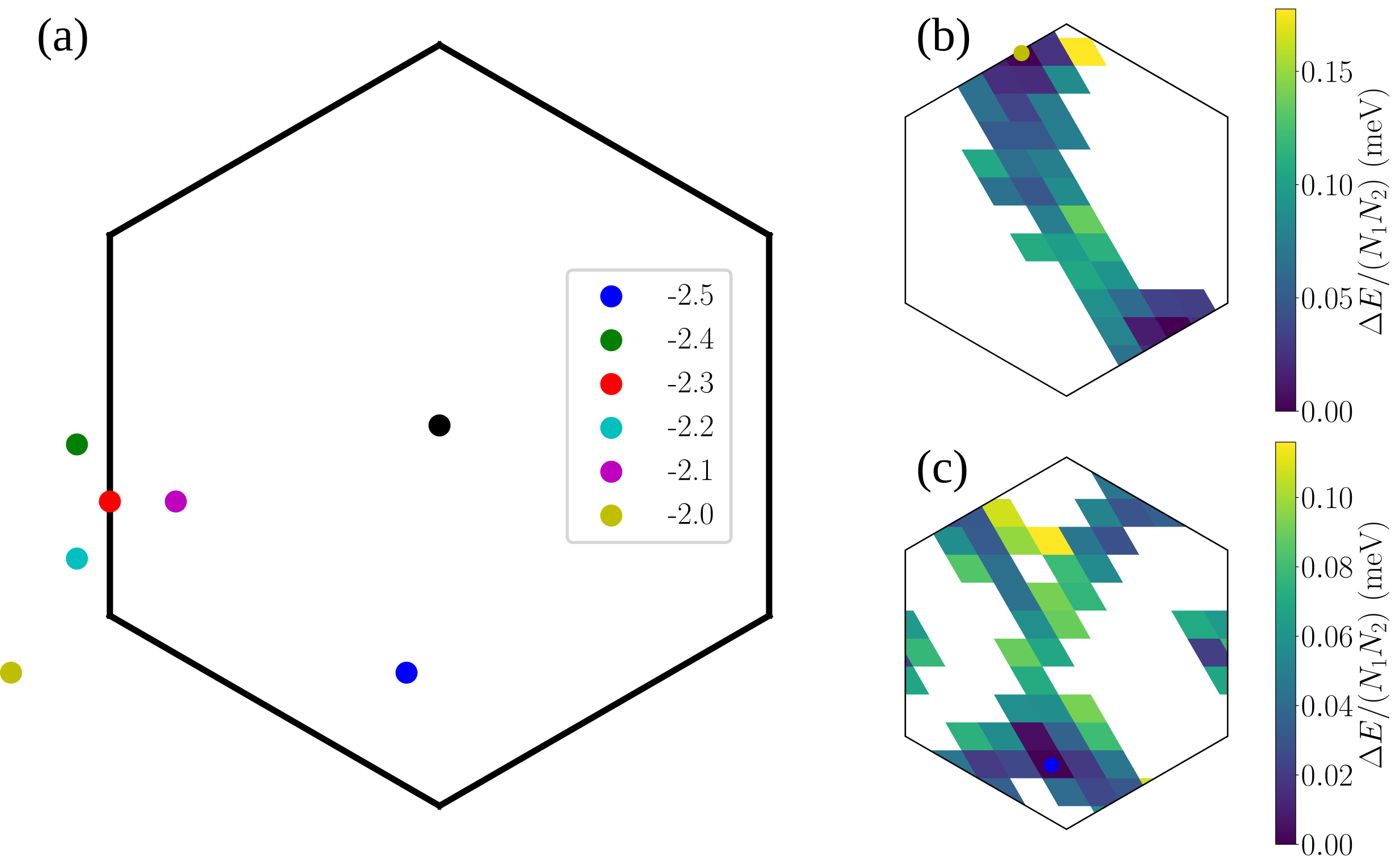}
    \caption{(a) Doping-dependent $\bm{q}_{\text{Kekul\'e}}$ vectors, found from $10 \times 10$ HF, of a system with $\epsilon = -0.12\%$ and $\varphi = 87^\circ$, consistent with the experimental parameters in Ref.~\cite{kim_imaging_2023}. We set $\Delta V = 0$ as displacement field in the single-gate geometry is weak, and use single-gate screened Coulomb interaction $V(q) = (e^2/2\epsilon_0\epsilon_rq)(1 - e^{-2qd})$ with screening length $d = 25$\,nm and relative permittivity $\epsilon_r = 20$. We take the middle-layer graphene to be rotated counter-clockwise, as in experiment but opposite to our usual convention. (b,c) $\bm{q}_{\text{Kekul\'e}}$-dependence of the energies of IKS states for $\nu = -2$ and $\nu = -2.5$ respectively, where $\Delta E = E(\bm{q}_{\text{Kekul\'e}}) - E_{\text{min}}$.}
    \label{fig:qiks}
\end{figure}

\textit{Comparison with Experiments.---}
Ref.~\cite{kim_imaging_2023} observes Kekul\'e patterns using STM in a TSTG device with heterostrain  of $-(0.12 \pm 0.04)\%$ and finds doping-dependent $\bm{q}_{\text{Kekul\'e}}$ vectors, which are equivalent to our $\bm{q}$ vectors up to choice of reference point, at fillings from $-2$ to $-2.5$. From our HF calculations with relative permittivity $\epsilon_r = 20$ (elsewhere in this paper, $\epsilon_r = 10$), at the experimental angle and magnitude of strain, we find that the ground states of  TSTG for experimental parameters % and fillings 
have IKS order, and we show the corresponding $\bm{q}_{\text{Kekul\'e}}$ in Figure.~\ref{fig:qiks}. We observe doping-dependent $\bm{q}_{\text{Kekul\'e}}$, in qualitative agreement with experiment. However, the values of $\bm{q}_{\text{Kekul\'e}}$ are highly sensitive to modelling details, and so we do not expect  precise quantitative agreement.

\textit{Concluding Remarks.---}Motivated by the experimental discovery of Kekul\'e spiral order in TSTG~\cite{kim_imaging_2023}, we have systematically studied its ground state phase diagram  under strain and interlayer potential. At zero interlayer potential, since the system decomposes into TBG and graphene sectors, the physics is similar to that of TBG, including the emergence of IKS order at finite strain. At finite interlayer potential, we find new effects such as the breaking of $C_2T$ symmetry and the formation of a commensurate Kekul\'e spiral state. Our prediction that the Kekul\'e spiral state at $\nu = 2$ breaks $C_2T$ symmetry and opens up a charge gap is in agreement with experimental results~\cite{shen_dirac_2023}. The \textit{commensurate} spiral state at zero strain is an addition to the family of Kekul\'e spiral states introduced by the discovery of the \textit{incommensurate} Kekul\'e spiral,  and hints at a broader relevance of KS order beyond strained samples. %
While the high interlayer potential required to stabilize the phase in TSTG necessitates a dual-gate device, precluding the direct detection of its Kekul\'e pattern using STM, the mechanism behind the formation of CKS order is more general, and could emerge in other moiré systems, especially in cases where $C_2$ symmetry is broken explicitly~\cite{kwan_strong-coupling_2023}. We have further investigated the normal (non-superconducting) state of TSTG at non-integer fillings. We establish a charge-transfer cascade mechanism between the TBG and graphene sectors in the absence of interlayer potential, and show that this is consistent with a picture where the two sectors are coupled by Hartree interactions and the TBG sector density of states tracks that of standalone TBG. Notably, we find that integer fillings of the TBG sector and `maximally insulating' densities need not coincide with overall integer filling, which has implications for experimentally mapping the phase diagram of TSTG.

\begin{acknowledgements}
\textit{Acknowledgements}.--- We thank Étienne Lantagne-Hurtubise for useful discussions about Ref.~\cite{kim_imaging_2023}, and akcnowledge support from EPSRC Grant EP/S020527/1, the European Research Council (ERC) under the European Union Horizon 2020 Research and Innovation Programme (Grant Agreement Nos. 804213-TMCS), and Leverhulme Trust International Professorship grant (number LIP-202-014). GW acknowledges funding from the University of Zurich postdoc grant FK-23-134. For the purpose of Open Access, the author has applied a CC-BY public copyright licence to any Author Accepted Manuscript version arising from this submission.

\end{acknowledgements}

\bibliography{refs.bib, refsTSTG.bib}

\setcounter{figure}{0}
\let\oldthefigure\thefigure
\renewcommand{\thefigure}{S\oldthefigure}

\setcounter{table}{0}
\renewcommand{\thetable}{S\arabic{table}}

\newpage
\clearpage

\begin{appendix}
\onecolumngrid
	\begin{center}\textbf{\large --- Supplementary Material ---}

\end{center}

\section{Strained BM Model for TSTG}

The unstrained BM model in one spin and valley for twisted trilayer graphene is given by~\cite{khalaf_magic_2019}
\begin{equation}
    H = \begin{pmatrix}
        -iv_0\bm{\sigma}_{\theta/2}\bm{\nabla_{\bm{r}}} + \Delta V/2 & T(\bm{r}) & 0\\
        T^\dagger(\bm{r}) & -iv_0\bm{\sigma}_{-\theta/2}\bm{\nabla_{\bm{r}}} & T^\dagger(\bm{r}) \\
        0 & T(\bm{r}) & -iv_0\bm{\sigma}_{\theta/2}\bm{\nabla_{\bm{r}}} - \Delta V /2
\end{pmatrix}
\end{equation} 
where $\bm{\sigma}_{\theta/2} = e^{-(i\theta/4) \sigma_z}(\sigma_x, \sigma_y)e^{(i\theta/4) \sigma_z}$, $T(\bm{r}) = \sum_{n=1}^{3} T_n e^{-i\bm{q}_n\bm{r}}$, and $\Delta V$ is the interlayer potential. The interlayer hopping matrices are given by $T_{n + 1} = w_{AA}\sigma_0 + w_{AB}(\sigma_x \cos(n \phi) + \sigma_y \sin(n \phi))$ where $\phi = 2\pi/3$ and the hopping vectors are given by $\bm{q}_1 = k_{\theta}(0, -1)$, $\bm{q}_{2,3} = k_{\theta} (\pm \sqrt{3}/2, 1/2)$, where $k_{\theta} = 2k_D\sin(\theta/2)$ and $k_D = 4\pi/3a_0$ is the Dirac momentum. Further, in all the calculations, we will ignore the Pauli twist in the kinetic energy term, i.e. replace $\bm{\sigma}_{\pm\theta/2}$ with simply $\bm{\sigma}$.

The effect of strain is incorporated in two ways. Firstly, moir\'e lattice vectors are modified, the details of which are developed and discussed in Ref.~\cite{parker_strain-induced_2021, bi_designing_2019}. Secondly, strain shifts the position of Dirac points through a vector potential~\cite{suzuura_phonons_2002}, which is given by 
\begin{equation}
    \bm{A} = \frac{\beta}{2a} (\epsilon_{xx} - \epsilon_{yy}, -2\epsilon_{xy})
\end{equation}
Here, the $2 \times 2$ strain tensor is given by
\begin{equation}
    S(\epsilon, \varphi) = R^{-1}(\varphi) \begin{pmatrix}
        - \epsilon & 0 \\
        0 & \nu \epsilon
    \end{pmatrix}R(\varphi)
\end{equation}
where $\epsilon$ is the magnitude of strain and $\varphi$ the strain angle, measured from the $x$-axis. $R(\varphi)$ is the usual rotation matrix. We will consider heterostrain that is equal in the top and bottom layers and opposite in the middle layer, with $\epsilon_{\text{top}} = \epsilon/2$, $\epsilon_{\text{middle}} = -\epsilon/2$ and $\epsilon_{\text{bottom}} = \epsilon/2$, and we will take $\varphi = 0$ for all layers unless otherwise noted. The strain vector potential is incorporated such that $-i\bm{\nabla_{\bm{r}}}  \rightarrow -i\bm{\nabla_{\bm{r}}} - \bm{A}$, where $\bm{A}$ is the strain vector potential for the corresponding layer. In the main text, we have considered $\epsilon > 0$, corresponding to compression of top and bottom layers and extension of the middle layer.

From Ref. \cite{khalaf_magic_2019}, we know that TSTG can be decomposed into a TBG sector and a graphene sector. The TBG sector has effective tunnelling strengths (i.e. $w_{AA}$ and $w_{BB}$) scaled by a factor of $\sqrt{2}$, and the corresponding magic angle is also scaled by a factor of $\sqrt{2}$. To reproduce the same single-particle band-structure, strain should also be scaled by a factor of $\sqrt{2}$. This is to say, there is a correspondence between a TBG system of angle $\theta$ and strain $\epsilon$ with the TBG sector of a TSTG system of angle $\sqrt{2}\theta$ and strain $\sqrt{2}\epsilon$. An intuitive understanding of the scaling of strain is the following: the size of the moir\'e BZ is proportional to $\theta$, while the shift of Dirac points due to strain vector potential $\bm{A}$ is proportional to $\epsilon$. As such, the ratio $\epsilon/\theta$ is a good measure of the effect of strain. This explains why we found phase transitions tending to happen at a larger strain in TSTG compared to TBG.

\section{Effect of Homostrain on Single-Particle Band Structure}

\begin{figure}
    \centering
    \includegraphics[width = 0.5\linewidth]{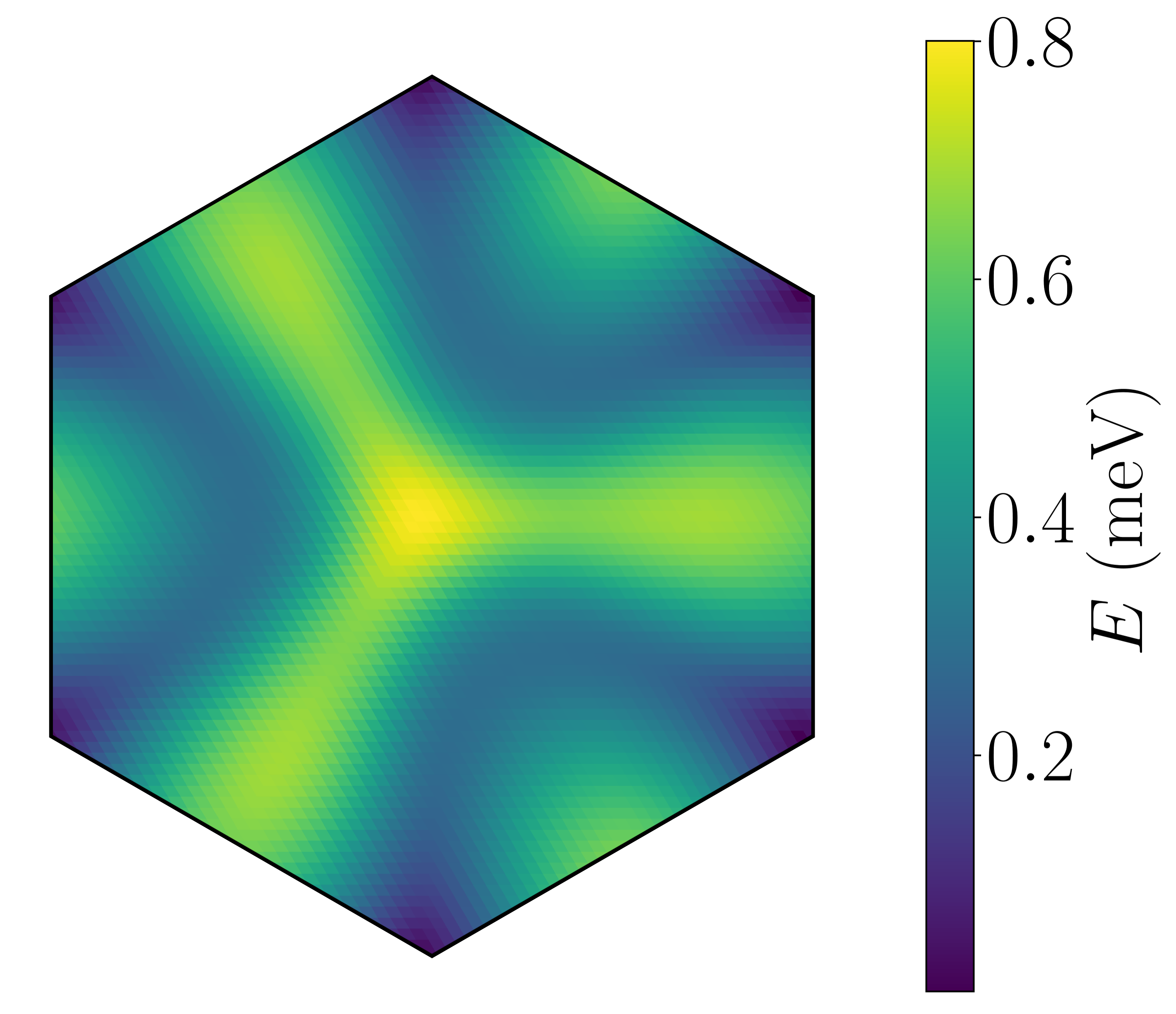}
    \caption{Single-particle central bands of a TSTG system under homostrain of $0.1\%$ ($\epsilon_1 = \epsilon_2 = \epsilon_3 = 0.1\%$). The band structure closely resembles that of unstrained system.}
    \label{fig:homo}
\end{figure}

In the main text, we have exclusively focused on heterostrain, where adjacent layers of graphene experience opposite strains. Homostrain, i.e. strain acting equally on all layers, has a much smaller effect on the band structure. In Figure \ref{fig:homo}, we show, within the continuum model, the central bands of the TBG sector of a TSTG system at $0.1\%$ homostrain. The band structure very closely resembles that of unstrained system, shown in Figure \ref{model}(d) in the main text. This is to be contrasted with heterostrained band structure in Figure \ref{model}(e).

\section{Hartree-Fock and Subtraction Scheme}

We perform Hartree-Fock (HF) iterations self-consistently. To keep the computation time manageable, we only retain a small number of active single-particle bands obtained from the BM model in previous section.  Unless otherwise stated, we keep 4 bands per spin/valley in our HF calculations.

Since the BM model band structures have already taken into account of some of the interaction, it is necessary to subtract away such effect. In practice, this amounts to replacing the density matrix $P$ with $P - P_0$ in computing the interaction term, where $P_0$ is the reference matrix. A few different methods exist, but in this work we have employed the average scheme: two central bands per spin/valley have occupation $1/2$, while bands above (below) the central bands are empty (filled). This is similar to the scheme employed in Ref. \cite{xie_twisted_2021}.

\section{Symmetry Properties of States Found in Phase Diagram}

In Table \ref{tab:symmetries}, we have summarized the symmetry properties of various states found in the main phase diagram.

\begin{table}[h!]
    \centering
    \begin{tabular}{c c c c c c c} 
    
    Phase & $\nu_s$ & Valley Pol. & $C_2T$ & $U_V(1)$ & $T$ & $T^\prime$ \\ \hline
    
    (I)KS(*) & 1 & $0^\dagger$ & * & \xmark & \checkmark$^\dagger$ & \xmark \\ 
    VP(*) & 1 & 1 & * & \checkmark & \xmark & \xmark \\
    K-IVC(*) & 0 & 0 & * & \xmark & \xmark & \checkmark \\  
    SM & 0 & 0 & \checkmark & \checkmark & \checkmark & \checkmark
    \end{tabular}
    \caption{The symmetries of various states that appear in the phase diagram. Since all states considered here can be reduced to colinear spin configurations with appropriate $\text{SU}(2)_K \times \text{SU}(2)_{K^\prime}$ rotations, we denote the filling of the spin sector where the state could occur as $\nu_s$. For states that break $U_V(1)$ symmetry, an appropriate $U_V(1)$ rotation is allowed before $C_2T$ symmetry is considered. Asterisk(*) labels variants of states that break $C_2T$ symmetry (e.g. VP preserves $C_2T$-symmetry but VP* breaks $C_2T$-symmetry). Note about notation: Kekul\'e spiral states could either break or conserve $C_2T$-symmetry. The $C_2T$-breaking version is termed KS*, which could either have commensurate or incommensurate $\bm{q}$. The $C_2T$-conserving version is called IKS, as there is no commensurate version of it in the phase diagrams. $\dagger$: At some regions, KS* states could have finite valley polarization and break time-reversal symmetry along the way. Refer to Figure 2 of the main text and the next section on phase boundaries for more discussion.}
    \label{tab:symmetries}
\end{table}

\section{Competition between K-IVC and T-IVC at $|\nu| = 2$}

Ref.~\cite{bultinck_ground_2020} predicts that the ground states of TBG at zero strain for fillings $|\nu| = 2$ are K-IVC states, and Ref.~\cite{parker_strain-induced_2021, kwan_kekule_2021} finds that this remains true at small enough strains. In this paper, we also find K-IVC to be the ground state for TSTG at small interlayer potential and strain at $|\nu| = 2$. However, for ultra-low strained TBG samples at fillings near $\nu = -2$, Ref.~\cite{nuckolls_quantum_2023} found a Kekul\'e density pattern, despite the fact that such a pattern is expected to vanish in the K-IVC from symmetry reasons~\cite{calugaru_spectroscopy_2022-1, hong_detecting_2022}. One candidate strong-coupling state that could manifest a Kekul\'e pattern is the T-IVC state, which was previously predicted to be energetically unfavorable~\cite{bultinck_ground_2020}. To resolve this contradiction, some of us proposed~\cite{kwan_electron-phonon_2023} that electron-phonon coupling could provide a mechanism that favors T-IVC over K-IVC at $|\nu| = 2$ based on self-consistent HF calculations~\cite{blason_local_2022}. Ref.~\cite{ingham_quadratic_2023} proposes an alternative mechanism based on renormalization group arguments.

While the aforementioned experimental measurements were made in a TBG system, TSTG with small interlayer potential resembles strongly a TBG system. As such, it is possible that the ground states of TSTG at $|\nu| = 2$ and at small interlayer potential and strain are actually T-IVC. 

\section{Phase Boundaries at $\nu = 2$}

\begin{figure}[h]
    \centering
    \includegraphics[width = 0.5\linewidth]{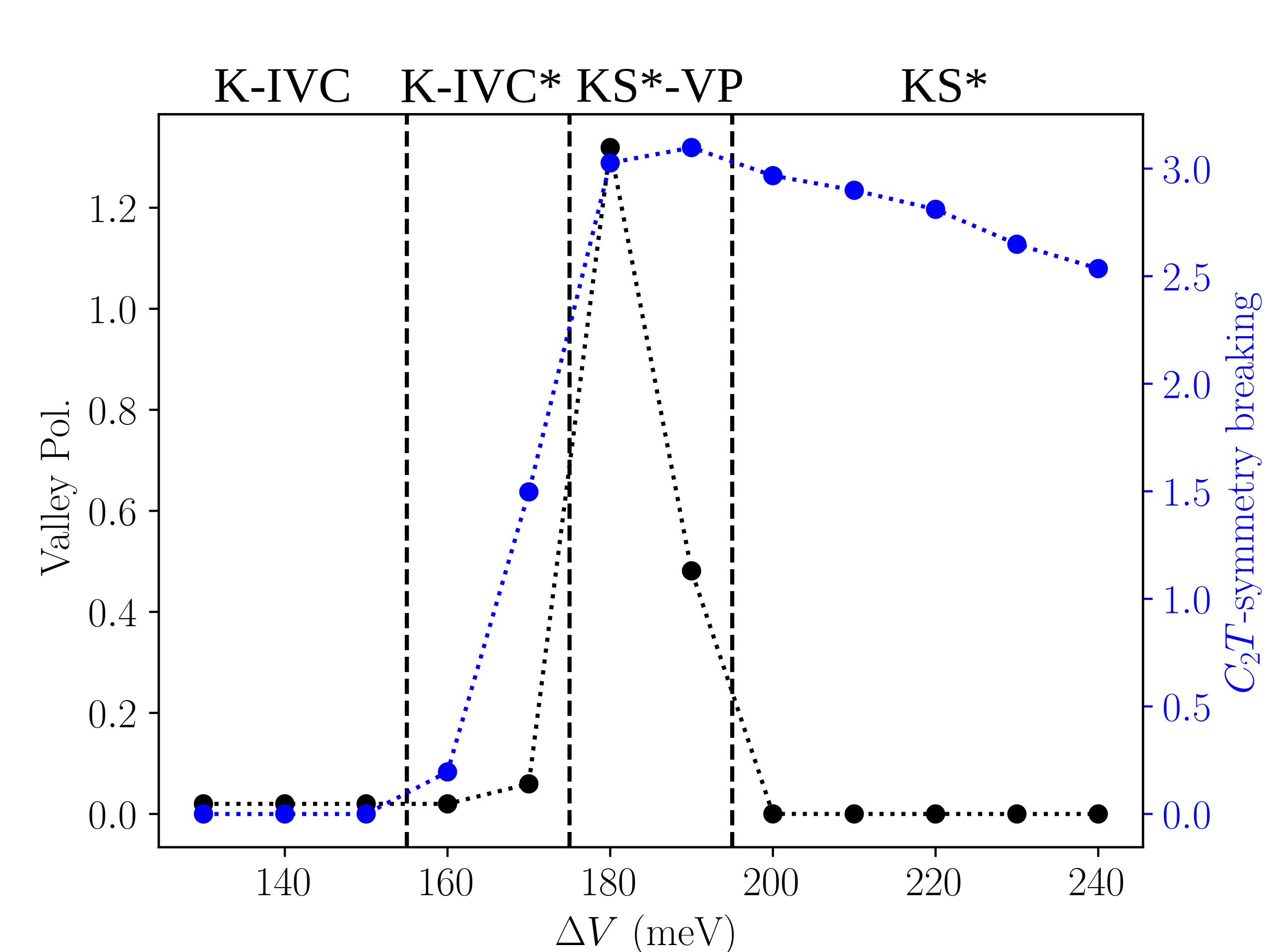}
    \caption{Properties and ground states at $\nu = 2$ without strain from $10 \times 10$ HF results. The phases in each parameter range is labelled above the graph.}
    \label{fig:nu2_transition}
\end{figure}

At $\nu = 2$, in the main phase diagram, we show the transition from K-IVC to KS* as we increase the interlayer potential at zero or small strain. In fact, we also observe intermediate phases of K-IVC* (which is the $C_2T$-symmetry breaking variant of K-IVC) and KS*-VP (which is a KS* state with finite valley polarization) in our HF numerics, as shown in Figure~\ref{fig:nu2_transition}. We believe that the order of transitions depends on the details of modelling, and in an actual experiment the intermediate phases may not be present.

\section{Momentum Space Properties of Commensurate Kekul\'e Spiral}

\begin{figure}[h]
    \centering
    \includegraphics[width=0.5\linewidth]{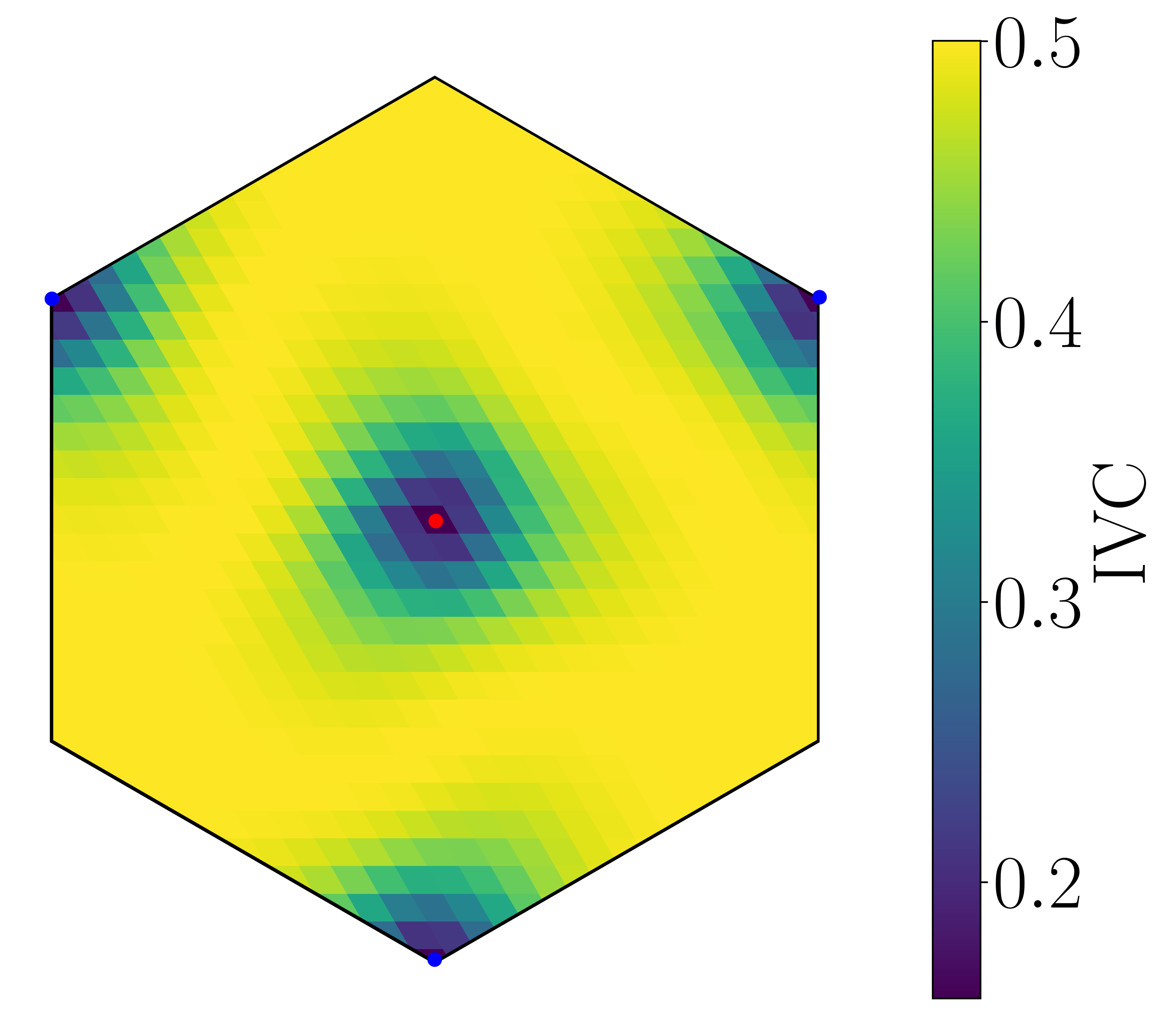}
    \caption{Momentum resolved intervalley coherence strength of CKS* from $24 \times 24$ self-consistent HF. The blue dots mark $K_M/\Gamma_M$ position (for $K$ and $K^\prime$ valleys), and the red dot marks $\Gamma_M/K^\prime_M$ position.}
    \label{fig:k_space_ivc}
\end{figure}

In Figure. \ref{fig:k_space_ivc}, we show the strength of intervalley coherence (IVC) in the moir\'e BZ for the CKS* state. We note that the regions where IVC strength is significantly reduced correspond to the $K_M$-point of $K$-valley or $K^\prime_M$-point of $K^\prime$-valley.

\section{Behaviours with Large Strains}
\begin{figure}[h]
    \centering
    \includegraphics[width=0.75\linewidth]{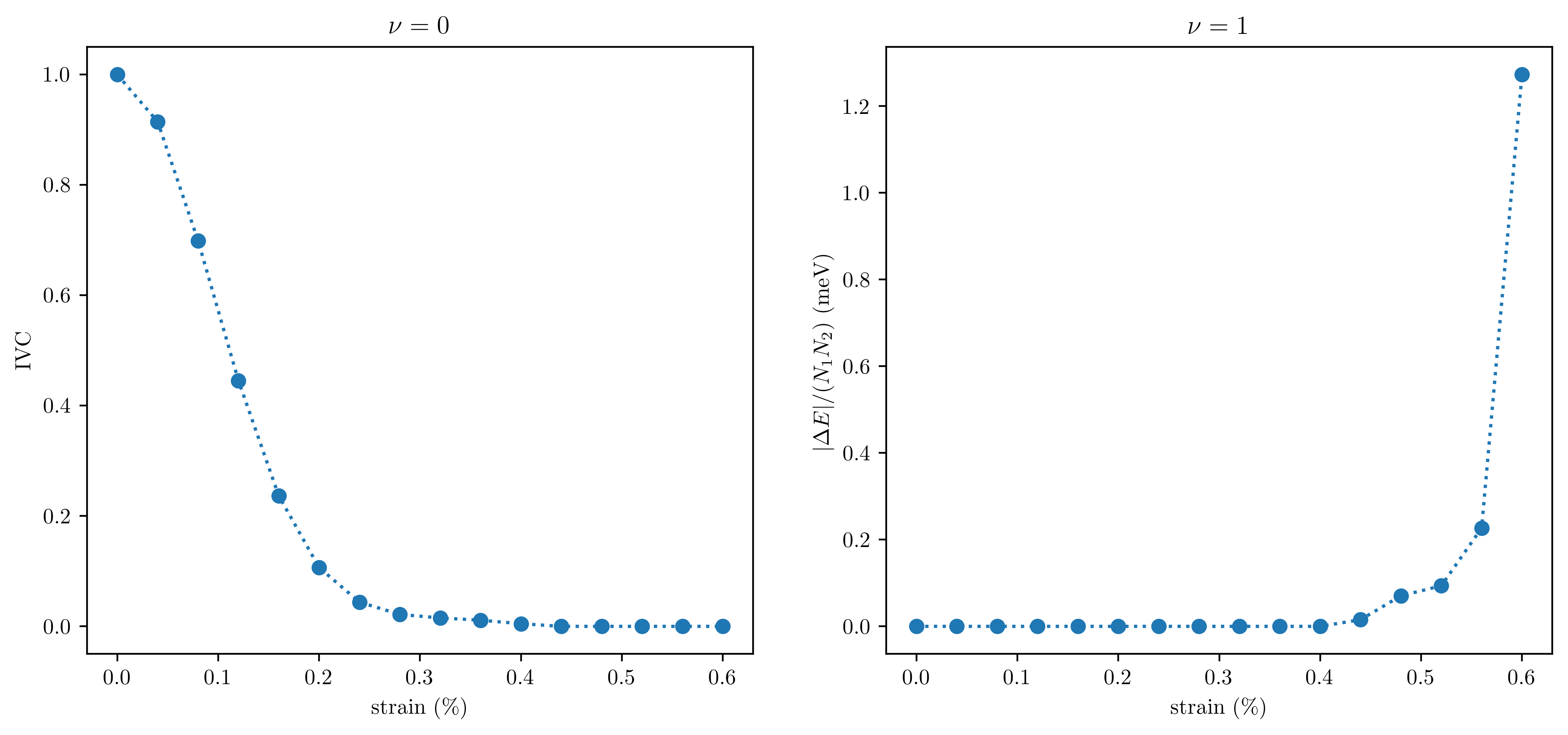}
    \caption{Systems with fillings $\nu = 0$ and $\nu = 1$ with strain up to $0.6\%$ and zero interlayer potential. The results are obtained from $12 \times 12$ self-consistent HF. On the left, ``IVC" is defined as $||P_{+-}||^2/(N_1N_2)$, the Frobenius norm squared of the intervalley part of the density matrix divided by the system size. On the right, $\Delta E$ is the energy difference between the $\bm{q} = 0$ state with the overall energy minimum. For the plot of $\nu = 0$, we choose to retain, instead of all bands closest to charge neutrality, 2 TBG bands and 2 graphene bands respectively closest to charge neutrality per spin and valley.}
    \label{fig:big_strain}
\end{figure}

Here we consider the behaviour with strains larger than those considered in the main text, up to $0.6\%$, at zero interlayer potential. As shown in Figure \ref{fig:big_strain}, at $\nu = 0$, there is a transition from K-IVC to symmetric semi-metal when going from the unstrained system to a system with large strain of about $0.5\%$. The Kekul\'e vector $\bm{q}$ remains zero wherever IVC is non-vanishing. At $\nu = 1$, there is onset of IKS from $0.4\%$ strain onwards.

\section{Comparison of Two Methods to Compute Charge Transfer}
\begin{figure}[h]
    \centering
    \includegraphics[width = 0.65\linewidth]{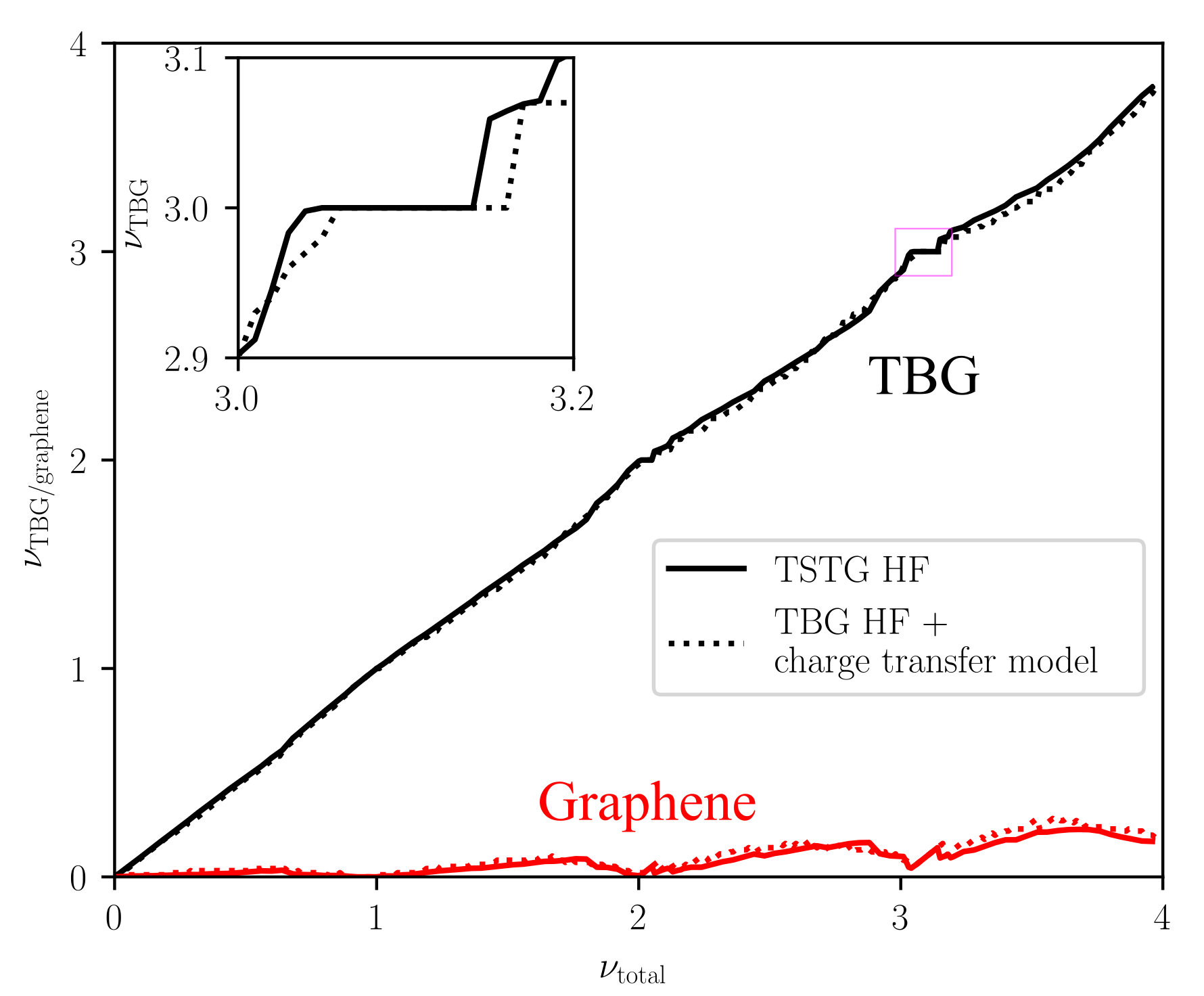}
    \caption{Comparison of fillings of TBG and graphene sectors obtained from two different methods. The results are from $10 \times 10$ HF calculations with strain of $0.3\%$ and zero interlayer potential.}
    \label{fig:comparison}
\end{figure}

As explained in the main text, the charge transfer between the TBG sector and graphene sector of a TSTG system can be understood in terms of matching the chemical potential of the two sectors. This allows the determination of the charge transfer with only Hartree-Fock calculations of TBG, where in addition the graphene sector can be modelled as simply non-interacting. The results of this method, labelled as `TBG HF', are compared with full TSTG HF calculation, shown in Figure \ref{fig:comparison}, and we obtain good agreement. The remaining discrepancy is likely the result of nonequivalent band cut-off implemented for TBG and TSTG calculations.

\section{Particle-Hole Symmetry Breaking and Negative Fillings}
As was shown in Ref.~\cite{calugaru_twisted_2021}, TSTG with interlayer potential but without strain has exact particle hole symmetry, which is given by a combined operator $m_zC_{2x}P$, where $P$ is the usual particle-hole symmetry, $m_z$ is mirror reflection in the $z$-plane, and $C_{2x}$ is $\pi$-rotation with respect to the $x$-axis. Now, applying strain would break $C_{2x}$ symmetry for Hamiltonian of the TBG sector ($\hat{H}_{\text{TBG}}$), and break the combined symmetry for $\hat{H}_{\text{TBG}}$ (strained graphene sector $\hat{H}_D$ also breaks $m_zC_{2x}P$ symmetry, though the reasoning is less trivial as it starts without $C_{2x}$ symmetry in the unstrained case). As such, there is no particle-hole symmetry for TSTG with both strain and interlayer potential~\footnote{In the case where no interlayer potential is applied, we can work around this by defining a particle-hole symmetry operator that is the direct sum of the correct operators on each sector, i.e. $P^\prime = (P)_{\text{TBG}} \oplus (C_{2x}P)_D$.}. 

It is easily seen that under $C_{2x}$ symmetry, for the TBG sector, strain is mapped from  $(\epsilon, \varphi)$ to $(-\epsilon, -\varphi)$, where $\epsilon$ is the magnitude of strain and $\varphi$ is the direction of strain measured from the $x$-axis. It can be shown that under the combined $m_zC_{2x}P$ the strain of the graphene sector transforms an the identical way, establishing a correspondence between system at $(\nu, \epsilon, \varphi)$ and system at $(-\nu, -\epsilon, -\varphi)$, where $\nu$ is the filling factor as usual.

\begin{figure}[h]
    \centering
    \includegraphics[width = 0.9\linewidth]{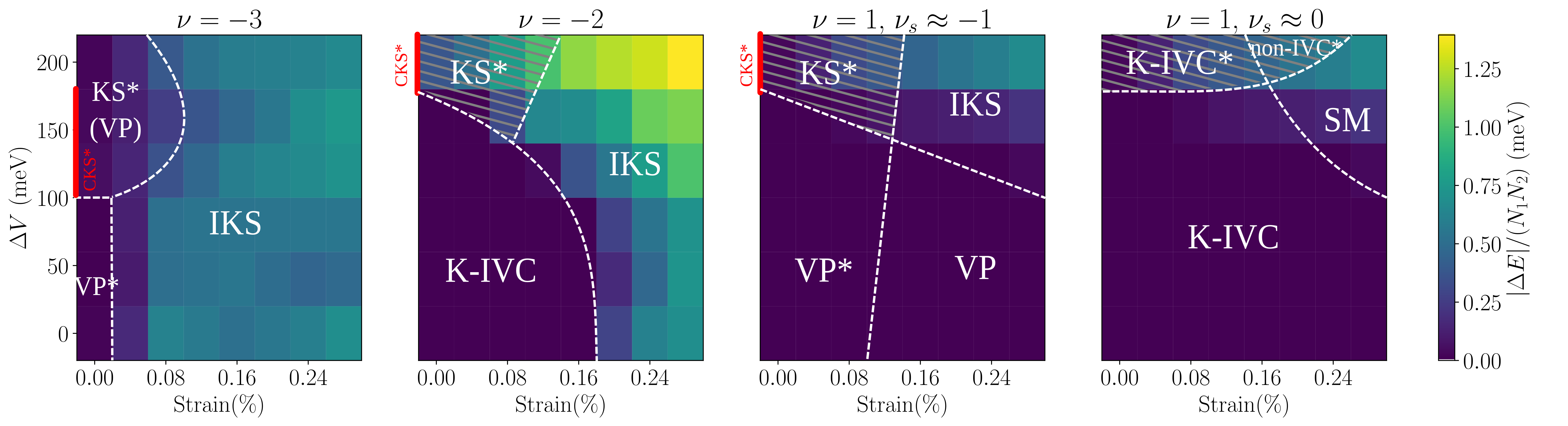}
    \caption{The phase diagram at negative fillings. The acronyms are the same as that used in the main text, with non-IVC* denotes a $C_2T$-breaking state that preserves other symmetries. Despite lack of exact particle-hole symmetry, there is little qualitative difference to the phase diagrams of positive fillings. Due to the rather coarse grid used in parameter space, the phase boundaries should be interpreted as a rough guide.}
    \label{fig:negative_filling}
\end{figure}

Lacking exact particle-hole symmetry, we computed the phase diagrams for the negative fillings as well, as shown in Figure~\ref{fig:negative_filling}. We see that there is no qualitative difference to that of the positive fillings. From the last paragraph, we understand that the results could also be applied to systems with positive fillings but negative strains.

\section{Translation Symmetry Breaking at High Interlayer Potential}

We perform self-consistent Hartree-Fock calculations allowing for all translation and spin symmetry breaking, and we observe the ground state to have complex spin textures at $\nu = 1$ with large interlayer potential ($> 200$ meV) and finite strain. The system has significant spatial fluctuations of spin densities, with much smaller total charge density modulation, which may be an artefact of our numerical methods. The spin patterns partially resemble spin spirals, but the spin rotations are not co-planar, and the results show strong finite size dependence. As this is not the main purpose of our work, the result remains inconclusive, but this could be the subject of future studies with careful control of finite size effects.

\begin{figure}
    \centering
    \includegraphics[width = 0.7\linewidth]{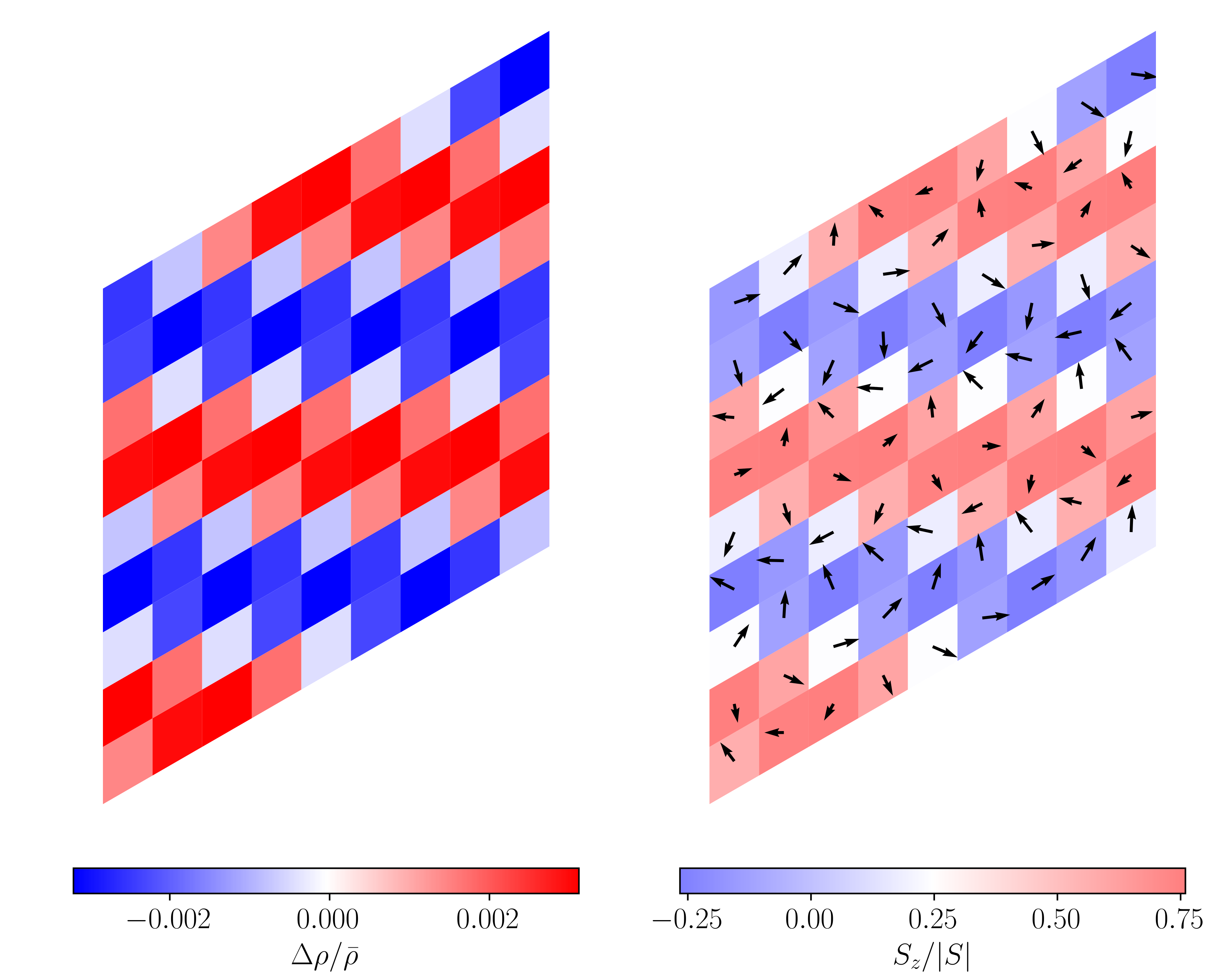}
    \caption{Spin texture of TSTG with $\epsilon = 0.16\%$ and $\Delta V = 240$~meV, obtained from $9 \times 9$ self-consistent HF, allowing for most general symmetry breakings. Since the system has $\text{SU}(2)_{K} \times \text{SU}(2)_{K^\prime}$ symmetry, allowing for independent spin rotations in each valley, we have only shown the spin texture of $K$-valley here.}
    \label{fig:enter-label}
\end{figure}

\section{Additional Results for Non-integer Fillings}

\begin{figure}[h]
    \centering
    \includegraphics[width = 0.7\linewidth]{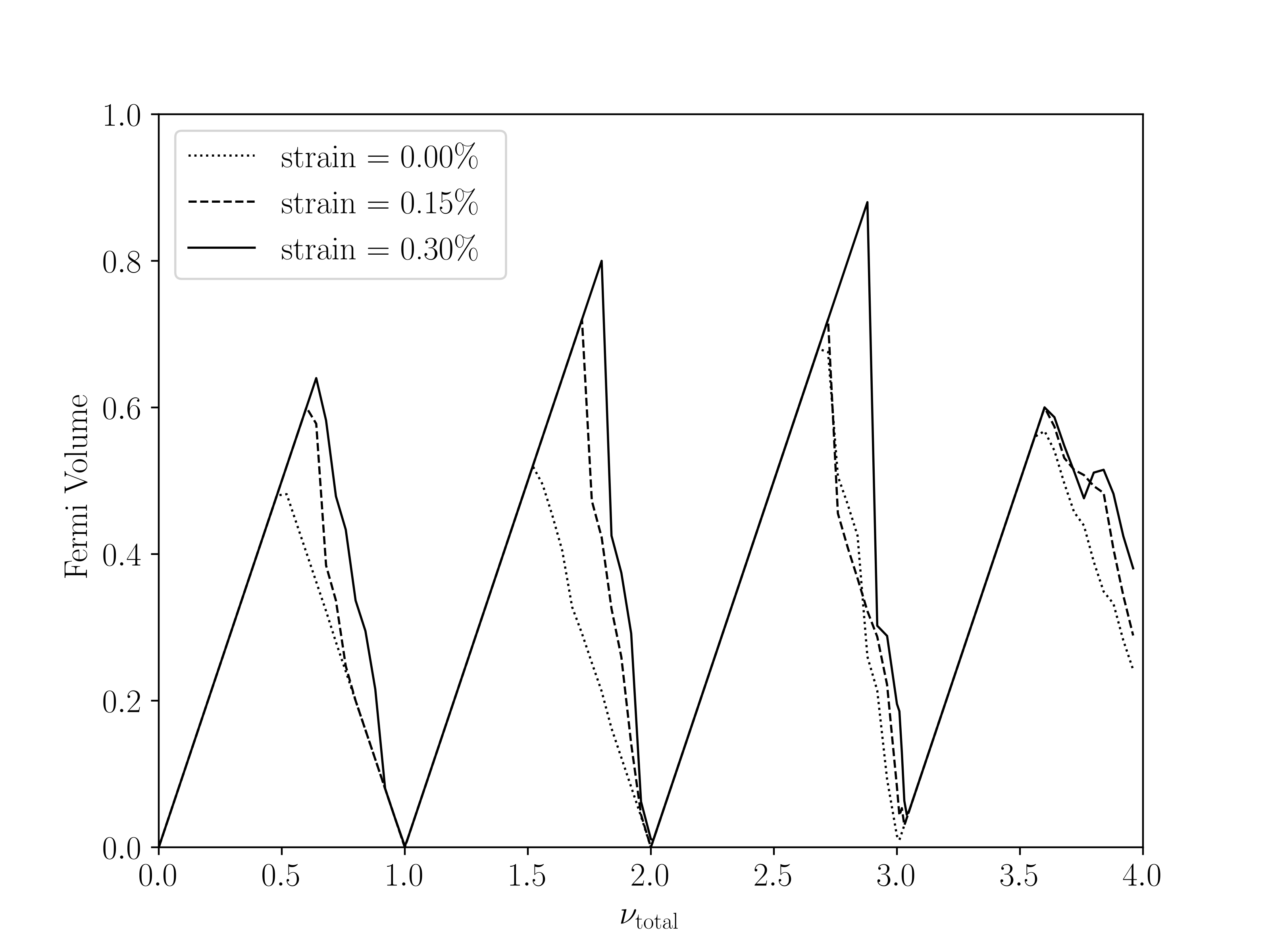}
    \caption{Total unsigned Fermi volume of TSTG system at zero interlayer potential, obtained from $10 \times 10$ HF.}
    \label{fig:enter-label}
\end{figure}

\end{appendix}

\end{document}